\UseRawInputEncoding
\documentclass{ws-ijmpa}
\usepackage[super,compress]{cite}
\usepackage{graphicx}
\begin{document}
\markboth{Z.-H. Weng}{Conserved quantities of vectorial magnitudes within the material media}

%
\catchline{}{}{}{}{}
%

\title{Conserved quantities of vectorial magnitudes within the material media
}

\author{Zi-Hua Weng
}

\address{School of Aerospace Engineering, Xiamen University, Xiamen, China
\\
College of Physical Science and Technology, Xiamen University, Xiamen, China
\\
xmuwzh@xmu.edu.cn}

%

\maketitle

\begin{history}
\received{Day Month Year}
\revised{Day Month Year}
\end{history}

\begin{abstract}
  By means of the rotational transformations of octonion coordinate systems, the paper aims to explore the physical properties of conserved quantities relevant to the vectorial magnitudes within the material media, revealing the simultaneity of some conserved quantities in the electromagnetic and gravitational fields. J. C. Maxwell first utilized the algebra of quaternions to describe the electromagnetic theory. The subsequent scholars studied the physical properties of electromagnetic and gravitational fields simultaneously, including the octonion linear momentum, angular momentum, torque and force within the material media. According to the algebra of octonions, the scalar parts of octonion physical quantities remain unchanged, while the vectorial parts may alter, in the rotational transformations of octonion coordinate systems. From the octonion characteristics, it is able to deduce a few conserved quantities related to the vectorial magnitudes in the octonion space $\mathbb{O}$, including the magnitudes of linear momentum, angular momentum, torque and force. Similarly, it is capable of inferring several conserved quantities relevant to the vectorial magnitudes in the transformed octonion space $\mathbb{O}_u$, including the magnitudes of electric current, magnetic moment and electric moment. Through the analysis and comparison, it is concluded that some conserved quantities, relevant to the vectorial magnitudes, are unable to be established simultaneously, from the point of view of the octonion spaces. This is helpful to deepen the further understanding of some conserved quantities related to the vectorial magnitudes.


\end{abstract}

\ccode{PACS numbers: 03.50.De, 02.20.-a, 04.50.-h, 11.30.Ly, 11.10.Kk}


\section{\label{sec:level1}Introduction}

May all conserved quantities be effective simultaneously? The ratio of the magnitude of electric current to that of linear momentum is a fixed value, is not it? Is the ratio of the magnitude of magnetic moment to that of angular momentum one fixed value? For a long time, these simple and important issues have puzzled and attracted the scholars. It was not until the emergence of the octonion field theory (short for the field theory described with the octonions) that these problems were answered partially. When the rotational transformation occurs in the octonion coordinate systems, the scalar part of one octonion physical quantity remains unchanged, although its vectorial part will vary. According to this octonion property, it is able to deduce the conserved quantities relevant to vectorial magnitudes, for the electromagnetic and gravitational fields within the material media. This extends the understanding of the physical properties of some conserved quantities related to the vectorial magnitudes.

At the beginning of the 20th century, W. Kaufmann first utilized the electromagnetic deflection method to measure the charge-to-mass ratio of the fast moving electron beams. And it is found that the charge-to-mass ratio in the electron beams would be decreased with the increase of velocities. This experimental result bewildered many scholars \cite{mpyle}. Subsequently, A. Bucherer and other scholars revealed that the charge-to-mass ratio does alter with the increase of velocities, in several higher precision experiments. Those experimental results show that the charge-to-mass ratio varies with the increase of velocity, providing an important experimental basis for the establishment of the special theory of relativity many years later. Nonetheless, the present experimental results show that there exist other factors to affect the charge-to-mass ratio.

In the existing field theories, there is a preconceived conjecture. The ratio, $\gamma$ , of the magnitude of magnetic moment to that of angular momentum may be one fixed value, for one charged particle. In the experimental measurement of electrons and muons \cite{babi,bennett}, several experimental results show that there is a prominent difference between the ratio $\gamma$ of muons \cite{albahri1,albahri2} and that of electrons. In order to explain the conjecture of the ratio $\gamma$ , a few scholars have to introduce another surmise. The latter assumes that the virtual particles in the vacuum affect the measurement of the ratio $\gamma$ of muons \cite{albahri3,bluhm1}. However, there is also an opposite viewpoint that the ratio $\gamma$ must not be a fixed value. Taking the ratio $\gamma$ as one fixed value is merely the subjective imagination and guess of several scholars, lacking sufficient experimental evidences. Strictly speaking, the gyro-magnetic ratio is not a fixed value either.

The above analysis shows that the existing field theories have some defects in the exploration of some conserved quantities related to the vectorial magnitudes, in the electromagnetic and gravitational fields. This faultiness restricts the scope of application of some conserved quantities relevant to the magnitudes of vectors.

1) Simultaneous establishment. The existing field theories take it for granted that all conserved quantities must be established simultaneously. However, this is only the subjective imagination of some scholars, while the subjective imagination does not have enough theoretical bases and experimental evidences. This point of view is unable to explain why the charge-to-mass ratio will change with various factors, on the premise that the conserved quantities are established simultaneously.

2) Conserved quantities. The magnitude of linear momentum and that of electric current are supposed to be both conserved quantities. In the existing field theories, there is even a point of view that the ratio of these two magnitudes is a fixed value. This point of view subjectively reckons that the conserved quantity related to the vectorial magnitudes is the same as that of scalars. Nevertheless in fact, there is a distinct essential difference between these two types of conserved quantities.

3) Simple relationship. The existing field theories conjecture that the magnitude of angular momentum and that of magnetic moment can be conserved quantities simultaneously. Sometimes the existing field theories even speculate that the ratio of these two magnitudes should be one fixed value. This desire for simplicity imagines the relationship between these two magnitudes to be too straightforward. This point of view even subjectively assumes that there is a simple linear relationship between these two magnitudes.

In stark contrast, the octonion field theory is capable of solving some problems of conserved quantities related to the vectorial magnitudes, exploring the premise of simultaneous establishment of the conserved quantities, for the gravitational and the electromagnetic fields within the material media. The application of octonions can study a few puzzles derived from the existing field theories, researching several physical properties of conserved quantities related to the vectorial magnitudes.

J. C. Maxwell first applied not only the algebra of quaternions but also the vector terminology to research the electromagnetic fields. After that, the subsequent scholars utilize the algebra of quaternions \cite{rawat,morita} and the algebra of octonions \cite{deleo,demir1} to explore the physical properties of electromagnetic fields \cite{mironov1}, gravitational fields \cite{tanisli1}, invariants\cite{tanisli2,tanisli3}, equilibrium equations, continuity equations, relativity \cite{moffat}, quantum mechanics \cite{gogberashvili,bernevig}, fluids \cite{demir2,mironov2}, plasmas \cite{demir3,demir4}, dyons \cite{chanyal3,chanyal4}, dark matter, astrophysical jets, strong nuclear fields \cite{chanyal2}, weak nuclear fields \cite{majid}, and black holes and others.

In the paper, the application of octonion field theory is capable of exploring some difficult problems of conserved quantities relevant to the vectorial magnitudes, in the electromagnetic and gravitational fields. This characteristic extends the applicable range of some conserved quantities related to the vectorial magnitudes.

1) Non-simultaneous establishment. According to the octonion field theory, all conserved quantities cannot be established simultaneously. Due to the inherent characteristics of octonions, some conserved quantities relevant to the vectorial magnitudes are in an octonion space, while the rest are in another octonion space. As a result, all conserved quantities are unable to be effective simultaneously. This point of view is able to explain why the charge-to-mass ratio will vary with various factors, in particular the velocities of particles.

2) Non-conserved quantities. The conserved quantities of vectorial magnitudes are distinct from those of scalar parts of octonion physical quantities. There are essential differences between the two types of conserved quantities. The magnitudes of linear momentum and electric current cannot be selected as the conserved quantities simultaneously. The ratio of these two magnitudes is unable to be a fixed value, in particular the electrons in the electron beams. There are not enough experimental results to support that the ratio of these two magnitudes is a fixed value.

3) Complicated relationship. For each charged particle, the magnitude of angular momentum and that of magnetic moment cannot be the conserved quantities simultaneously. According to the octonion field theory, the ratio of these two magnitudes will not be a conserved quantity, in particular the electrons and muons \cite{bluhm2}. Apparently the relationship between the magnitudes of angular momentum and magnetic moment is quite complicated. There is no simple linear relationship between the two physical quantities.

In the paper, the scalar part of each octonion physical quantity keeps the same, while the vectorial part will transfer, when the rotational transformation occurs in the octonion coordinate systems. By means of this property of octonions, one can obtain some conserved quantities related to the vectorial magnitudes. In the octonion space $\mathbb{O}$ , it is able to achieve a few conserved quantities, relevant to the magnitudes of linear momentum, angular momentum, torque and force. Due to the limitation of simultaneity, these conserved quantities related to the vectorial magnitudes are unable to be established simultaneously, in the octonion space $\mathbb{O}$. Similarly, it is capable of obtaining several conserved quantities, relevant to the magnitudes of electric current, magnetic moment and electric moment, in the transformed octonion space $\mathbb{O}_u$ (in Section 3). Either these conserved quantities relevant to the vectorial magnitudes are unable to be established simultaneously, because of the limitation of simultaneity in the transformed octonion space $\mathbb{O}_u$ . These results are helpful to further understand the physical properties of the conserved quantities related to the vectorial magnitudes.

\begin{table}[h]
\tbl{Some octonion composite physical quantities have considered the contributions of material media, in the octonion spaces for the electromagnetic and gravitational fields.}
{\begin{tabular}{@{}ll@{}}
\hline\hline
physical quantity                     &  equation                                                                                                               \\
\hline
composite field source                & $\mu \mathbb{S}^+ = - ( i \mathbb{F}^+ / v_0 + \lozenge )^\ast \circ \mathbb{F}^+$                                      \\
composite linear momentum             & $\mathbb{P}^+ = \mu \mathbb{S}^+ / \mu_g$                                                                               \\
composite angular momentum            & $\mathbb{L}^+ = ( \mathbb{R} + k_{rx} \mathbb{X} )^\times \circ \mathbb{P}^+$                                           \\
composite torque                      & $\mathbb{W}^+ = - v_0 ( i \mathbb{F}^+ / v_0 + \lozenge ) \circ \{ ( i \mathbb{V}^\times / v_0 ) \circ \mathbb{L}^+ \}$ \\
composite force                       & $\mathbb{N}^+ = - ( i \mathbb{F}^+ / v_0 + \lozenge ) \circ \{ ( i \mathbb{V}^\times / v_0 ) \circ \mathbb{W}^+ \}$     \\
\hline\hline
\end{tabular}}
\end{table}

\section{\label{sec:level1}Material media}

R. Descartes considered that the space is the extension of substance. The Cartesian thought has been developed today, that is, any fundamental space is the extension of one fundamental field. The fundamental fields involve the gravitational field and electromagnetic field. Each fundamental field possesses one fundamental space. Each fundamental space is the quaternion space. Two independent quaternion spaces can be combined together to become one octonion space.

After W. R. Hamilton invented the quaternions, J. T. Graves and A. Cayley discovered the octonions, respectively. This octonion is called the classical octonion. It is capable of applying the classical octonions to explore the physical properties of electromagnetic and gravitational fields simultaneously, in this paper.

In the octonion space $\mathbb{O}$ for the electromagnetic and gravitational fields, $\textbf{i}_j$ and $\textbf{I}_j$ are the basis vectors, with $\textbf{I}_k = \textbf{i}_k \circ \textbf{I}_0$ . The octonion radius vector is, $\mathbb{R} = ( i \textbf{i}_0 r_0 + \Sigma \textbf{i}_k r_k ) + k_{eg} ( i \textbf{I}_0 R_0 + \Sigma \textbf{I}_k R_k )$. The octonion velocity is, $\mathbb{V} = ( i \textbf{i}_0 v_0 + \Sigma \textbf{i}_k v_k ) + k_{eg} ( i \textbf{I}_0 V_0 + \Sigma \textbf{I}_k V_k )$. Herein $\textbf{i}_0 = 1$. $\textbf{i}_k^2 = -1$. $\textbf{I}_j^2 = -1$. $r_j$ and $R_j$ are the coordinate values, while $v_j$ and $V_j$ are the speed values. $k_{eg}$ is the coefficient, to satisfy the needs of dimensional homogeneity \cite{weng1}. $r_j$ , $v_j$ , $R_j$ , and $V_j$ are all real. $r_0 = v_0 t$. $v_0$ is the speed of light, while $t$ is the time. $\circ$ denotes the octonion multiplication. $i$ is the imaginary unit. $j = 0, 1, 2, 3$. $k = 1, 2, 3$.

In the octonion space without considering the contribution of the material media, the octonion field strength is $\mathbb{F}$, the octonion field source is $\mathbb{S}$ , and the octonion linear momentum is $\mathbb{P}$ . Next, it is able to successively define the octonion angular momentum $\mathbb{L}$ , torque $\mathbb{W}$ , and force $\mathbb{N}$ and others (see Ref.[28]), from the octonion linear momentum and radius vector and others. Herein $\mu$ is the coefficient. $\mu_g$ is the gravitational constant, while $\mu_e$ is the electromagnetic constant. $k_{eg}^2 = \mu_g / \mu_e$ . The quaternion operator is, $\lozenge = i \textbf{i}_0 \partial_0 + \Sigma \textbf{i}_k \partial_k$ , with $\partial_j = \partial / \partial r_j$ . $\ast$ denotes the octonion conjugate.

In order to achieve the physical quantities within the material media, the octonion field strength and angular momentum can be combined into the octonion composite field strength, $\mathbb{F}^+ = \mathbb{F} + k_{fl} \mathbb{L}$ . In other words, the term $\mathbb{F}^+$ is the octonion field strength within the material media, while the term $\mathbb{F}$ is the octonion field strength in a vacuum. Herein $k_{fl} = - \mu_g$ is the coefficient, to meet the requirements of dimensional homogeneity (see Ref.[28]).

In the octonion space considering the contribution of the material media, the octonion composite field strength is $\mathbb{F}^+$ , the octonion composite field source is $\mathbb{S}^+$, and the octonion composite linear momentum is $\mathbb{P}^+$ . Subsequently, it is capable of defining successively the octonion composite angular momentum $\mathbb{L}^+$ , composite torque $\mathbb{W}^+$ and composite force $\mathbb{N}^+$, from the physical quantities $\mathbb{P}^+$ and $\mathbb{R}$ and others (Table 1).

In order to explore the physical properties of electromagnetic media and gravitational media, it is necessary to introduce the octonion composite physical quantities. The octonion composite physical quantities can be utilized to research several significant physical properties of electromagnetic and gravitational fields within the material media. Therefore it is capable of studying the electromagnetic equations within the electromagnetic media, the gravitational equations within the gravitational media, and a few invariants within the material media.

\begin{table}[h]
\tbl{Two conserved quantities, relevant to the vectorial magnitudes of the octonion linear momentum, $\mathbb{P}^+$, for the electromagnetic and gravitational fields within the material media.}
{\begin{tabular}{@{}llll@{}}
\hline\hline
physical quantity               &  magnitude                      &  unit vector                   &  octonion space                \\
\hline
$\mathbb{P}^+$                  &  $p^+$                          &  $\textbf{i}_{\{p+\}}$           &  $\mathbb{O}$                  \\
                                &  $P^+$                          &  $\textbf{I}_{\{P+\}}$           &  $\mathbb{O}_u$                \\
\hline\hline
\end{tabular}}
\end{table}

\section{\label{sec:level1}Composite linear momentum}

In terms of the quaternion spaces, the coordinate transformation of one uniform relative motion between two coordinate systems, in the two-dimensional spaces, is only a rotational transformation of the quaternion spaces in point of fact. Further this type of method can also be extended to the relevant circumstances in the octonion spaces, in particular the conserved quantities relevant to the scalar parts of octonions \cite{weng2}.

The conserved quantities related to the vectorial magnitudes are applied to a great extent in the physical sciences. However, compared with the conserved quantities related to the scalar parts of octonion physical quantities \cite{weng3}, there are more restrictions on the applicable conditions of these conserved quantities, related to the vectorial magnitudes of octonion physical quantities, as a result their scope of application is quite narrow.

By means of two unit vectors of the composite linear momentum in the octonion spaces, it is able to achieve two conserved quantities related to the vectorial magnitudes, under certain correlation conditions. a) The magnitude of linear momentum is a conserved quantity; b) The magnitude of electric current is a conserved quantity (Table 2).

The octonion composite linear momentum can be defined as, $\mathbb{P}^+ = \mu \mathbb{S}^+ / \mu_g$ , within the material media. And that the definition can be rewritten as,
\begin{eqnarray}
& \mathbb{P}^+ = ( i p_0^+ + \textbf{p}^+ ) + k_{eg} ( i \textbf{P}_0^+ + \textbf{P}^+ ) ~,
\nonumber
\end{eqnarray}
or
\begin{eqnarray}
& \mathbb{P}^+ = ( i p_0^+ + p^+ \textbf{i}_{\{p+\}} ) + k_{eg} ( i \textbf{P}_0^+ + P^+ \textbf{I}_{\{P+\}}  ) ~,
\end{eqnarray}
where $\textbf{i}_{\{p+\}}$ and $\textbf{I}_{\{P+\}}$ both are the unit vectors. $\textbf{p}^+ = p^+ \textbf{i}_{\{p+\}}$ . $\textbf{P}^+ = P^+ \textbf{I}_{\{P+\}}$. $\textbf{p}^+ = \Sigma p_k^+ \textbf{i}_k$ . $\textbf{P}^+ = \Sigma P_k^+ \textbf{I}_k$. $\textbf{P}_0^+ = P_0^+ \textbf{I}_0$ . $\textbf{p}^+$ is the linear momentum within the material media. $\textbf{P}^+$ is proportional to the electric current within the material media. $\textbf{P}_0^+$ is proportional to the electric charge within the material media. $p_0^+$ , $p_k^+$, $P_0^+$ , and $P_k^+$ are all real.

In the octonion spaces, we had discussed the conserved quantities relevant to the scalar parts, $p_0^+$ and $P_0^+$ , of the octonion composite linear momentum (see Ref.[30]). In the paper, it is going to study the conserved quantities related to the vectorial parts, $\textbf{p}^+$ and $\textbf{P}^+$ , of the octonion composite linear momentum.

1) In the octonion space $\mathbb{O}$ , if you multiply the unit vector, $\textbf{i}_{\{p+\}}$ , by Eq.(1) from the left, you will get,
\begin{eqnarray}
& \textbf{i}_{\{p+\}} \circ \mathbb{P}^+ = ( i p_0^+ \textbf{i}_{\{p+\}} - p^+ ) + k_{eg} \textbf{i}_{\{p+\}} \circ ( i \textbf{P}_0^+ + P^+ \textbf{I}_{\{P+\}}  ) ~,
\end{eqnarray}
where $( \textbf{i}_{\{p+\}} )^2 = -1$.

The scalar part, $( - p^+ )$, in the above remains unchanged, in the rotational transformations of the octonion coordinate systems. In other words, the magnitude of linear momentum will keep the same in the octonion rotational transformations. Apparently, the octonion composite linear momentum, $( \textbf{i}_{\{p+\}} \circ \mathbb{P}^+ )$ , in Eq.(2) is distinct from $\mathbb{P}^+$ in Eq.(1). Consequently, the scalar parts of these two octonion composite linear momenta are unable to be established simultaneously. If the magnitude of linear momentum is taken as one conserved quantity, its application range will be limited to some extent.

For the vectorial part, $\textbf{Z}_p^+ = \textbf{p}^+ + k_{eg} ( i \textbf{P}_0^+ + \textbf{P}^+ )$ , of $\mathbb{P}^+$, if you multiply the unit vector, $\textbf{e}_{\{p+\}}$ , by $\textbf{Z}_p^+$ from the left, the scalar part, $\textbf{e}_{\{p+\}} \circ \textbf{Z}_p^+$ , will remain unchanged, in the rotational transformations of the octonion coordinate systems. Herein $\textbf{e}_{\{p+\}}$ is the unit vector of vectorial part $\textbf{Z}_p^+$ .

2) In the transformed octonion space $\mathbb{O}_u$ , if we multiply the unit vector, $i \textbf{I}_{\{P+\}}$, by Eq.(1) from the left, we shall attain,
\begin{eqnarray}
& i \textbf{I}_{\{P+\}} \circ \mathbb{P}^+ = \textbf{I}_{\{P+\}} \circ ( - p_0^+ + i p^+ \textbf{i}_{\{p+\}} ) - k_{eg} ( \textbf{I}_{\{P+\}} \circ \textbf{P}_0^+ + i P^+ ) ~,
\end{eqnarray}
where $( \textbf{I}_{\{P+\}} )^2 = -1$.

Similarly, the scalar part, $( - P^+ )$, in the above keeps the same, in the rotational transformations of the octonion coordinate systems. That is, the magnitude of electric current will remain unchanged in the octonion rotational transformations. Obviously, the octonion composite linear momentum, $( i \textbf{I}_{\{P+\}} \circ \mathbb{P}^+ )$ , in Eq.(3) is different from not only the term $\mathbb{P}^+$ in Eq.(1), but also the term $( \textbf{i}_{\{p+\}} \circ \mathbb{P}^+ )$ in Eq.(2). As a result, the scalar parts of these three octonion composite linear momenta are unable to be effective simultaneously. Next, if the magnitude of electric current is chosen as one conserved quantity, its application range must be confined to some extent.

For the vectorial part, $\textbf{Z}_P^+$ , of transformed octonion composite linear momentum $i \textbf{I}_0 \circ \mathbb{P}^+$, if you multiply the unit vector, $\textbf{e}_{\{P+\}}$ , by $\textbf{Z}_P^+$ from the left, the scalar part, $\textbf{e}_{\{P+\}} \circ \textbf{Z}_P^+$ , will keep the same, in the rotational transformations of the octonion coordinate systems. Herein $\textbf{e}_{\{P+\}}$ is the unit vector of vectorial part $\textbf{Z}_P^+$ .

Since the transformed octonion space $\mathbb{O}_u$ is independent of the octonion space $\mathbb{O}$ , the magnitude of the electric current and that of the linear momentum cannot be simultaneously selected as conserved quantities.

The above method can be extended to some of other conserved quantities, related to the vectorial magnitudes of octonion physical quantities, in particular the octonion composite angular momentum.

\begin{table}[h]
\tbl{Four conserved quantities, relevant to the vectorial magnitudes of the octonion angular momentum, $\mathbb{L}^+$, for the electromagnetic and gravitational fields within the material media.}
{\begin{tabular}{@{}llll@{}}
\hline\hline
physical quantity               &  magnitude                      &  unit vector                   &  octonion space                \\
\hline
$\mathbb{L}^+$                  &  $L_1^{i+}$                     &  $\textbf{i}_{\{L1i+\}}$         &  $\mathbb{O}$                  \\
                                &  $L_1^+$                        &  $\textbf{i}_{\{L1+\}}$          &  $\mathbb{O}$                  \\
                                &  $L_2^{i+}$                     &  $\textbf{I}_{\{L2i+\}}$         &  $\mathbb{O}_u$                \\
                                &  $L_2^+$                        &  $\textbf{I}_{\{L2+\}}$          &  $\mathbb{O}_u$                \\
\hline\hline
\end{tabular}}
\end{table}

\section{\label{sec:level1}Composite angular momentum}

Making use of four unit vectors of the composite angular momentum in the octonion spaces, it is capable of achieving four conserved quantities related to the vectorial magnitudes, under certain correlation conditions. a) The magnitude of angular momentum is a conserved quantity; b) The magnitude of electric moment is a conserved quantity; c) The magnitude of magnetic moment is a conserved quantity (Table 3).

The octonion composite angular momentum can be written as, $\mathbb{L}^+ = ( \mathbb{R} + k_{rx} \mathbb{X} )^\times \circ \mathbb{P}^+$ , within the material media. And that the definition can be expanded as,
\begin{eqnarray}
& \mathbb{L}^+ = ( L_{10}^+ + i \textbf{L}_1^{i+} + \textbf{L}_1^+ ) + k_{eg} ( \textbf{L}_{20}^+ + i \textbf{L}_2^{i+} + \textbf{L}_2^+ ) ~,
\end{eqnarray}
where $\textbf{L}_1^+ = \Sigma L_{1k}^+ \textbf{i}_k$ , $\textbf{L}_1^{i+} = \Sigma L_{1k}^{i+} \textbf{i}_k$ . $\textbf{L}_2^+ = \Sigma L_{2k}^+ \textbf{I}_k$ , $\textbf{L}_2^{i+} = \Sigma L_{2k}^{i+} \textbf{I}_k$ . $\textbf{L}_{20}^+ = L_{20}^+ \textbf{I}_0$ . $\textbf{L}_1^+$ is the angular momentum within the material media. $\textbf{L}_2^{i+}$ is the electric moment within the material media, while $\textbf{L}_2^+$ is the magnetic moment within the material media. $\mathbb{X}$ is the integrating function of octonion field potential. $\times$ denotes the complex conjugate. $k_{rx} = 1 / v_0$ is the coefficient, to meet the requirements of dimensional homogeneity. $L_{1j}^+$ , $L_{2j}^+$ , $L_{1k}^{i+}$ , and $L_{2k}^{i+}$ are all real.

The above can be further rewritten as,
\begin{eqnarray}
\mathbb{L}^+ = && ( L_{10}^+ + i L_1^{i+} \textbf{i}_{\{L1i+\}} + L_1^+ \textbf{i}_{\{L1+\}} )
\nonumber
\\
&&
+ k_{eg} ( \textbf{L}_{20}^+ + i L_2^{i+} \textbf{I}_{\{L2i+\}} + L_2^+ \textbf{I}_{\{L2+\}} ) ~,
\end{eqnarray}
where each of $\textbf{i}_{\{L1+\}}$ , $\textbf{i}_{\{L1i+\}}$ , $\textbf{I}_{\{L2+\}}$ , and $\textbf{I}_{\{L2i+\}}$ is one unit vector. $\textbf{L}_1^{i+} = L_1^{i+} \textbf{i}_{\{L1i+\}}$. $\textbf{L}_1^+ = L_1^+ \textbf{i}_{\{L1+\}}$. $\textbf{L}_2^{i+} = L_2^{i+} \textbf{I}_{\{L2i+\}}$. $\textbf{L}_2^+ = L_2^+ \textbf{I}_{\{L2+\}}$.

In the octonion spaces, the author had studied the conserved quantities relevant to the scalar parts, $L_{10}^+$ and $L_{20}^+$ , of the octonion composite angular momentum (see Ref.[30]). In the paper, we shall explore the conserved quantities related to the vectorial parts, $\textbf{L}_1^{i+}$ , $\textbf{L}_1^+$ , $\textbf{L}_2^{i+}$ and $\textbf{L}_2^+$ , of the octonion composite angular momentum.

1) In the octonion space $\mathbb{O}$ , if we multiply the unit vector, $\textbf{i}_{\{L1+\}}$ , by Eq.(5) from the left, we shall get,
\begin{eqnarray}
\textbf{i}_{\{L1+\}} \circ \mathbb{L}^+ = && ( L_{10}^+ \textbf{i}_{\{L1+\}} + i L_1^{i+} \textbf{i}_{\{L1+\}} \circ \textbf{i}_{\{L1i+\}} - L_1^+ )
\nonumber
\\
&& + k_{eg} \textbf{i}_{\{L1+\}} \circ ( \textbf{L}_{20}^+ + i L_2^{i+} \textbf{I}_{\{L2i+\}} + L_2^+ \textbf{I}_{\{L2+\}} ) ~,
\end{eqnarray}
where $( \textbf{i}_{\{L1+\}} )^2 = -1$.

The scalar part, $( - L_1^+ )$, in the above remains unchanged, in the rotational transformations of the octonion coordinate systems. In other words, the magnitude of angular momentum will keep the same in the octonion rotational transformations. Significantly, the octonion composite angular momentum, $( \textbf{i}_{\{L1+\}} \circ \mathbb{L}^+ )$, in Eq.(6) is distinct from $\mathbb{L}^+$ in Eq.(5). Therefore the scalar parts of these two octonion composite angular momenta are incompatible simultaneously. If the magnitude of angular momentum is considered as one conserved quantity, its application range has to be limited to some extent.

In the octonion space $\mathbb{O}$ , if you multiply the unit vector, $\textbf{i}_{\{L1i+\}}$ , by Eq.(5) from the left, you will obtain,
\begin{eqnarray}
\textbf{i}_{\{L1i+\}} \circ \mathbb{L}^+ = && ( L_{10}^+ \textbf{i}_{\{L1i+\}} - i L_1^{i+} + L_1^+ \textbf{i}_{\{L1i+\}} \circ \textbf{i}_{\{L1+\}})
\nonumber
\\
&& + k_{eg} \textbf{i}_{\{L1i+\}} \circ ( \textbf{L}_{20}^+ + i L_2^{i+} \textbf{I}_{\{L2i+\}} + L_2^+ \textbf{I}_{\{L2+\}} ) ~,
\end{eqnarray}
where $( \textbf{i}_{\{L1i+\}} )^2 = -1$.

In the rotational transformations of the octonion coordinate systems, the scalar part, $( - i L_1^{i+} )$, in the above keeps the same. In other words, the term $L_1^{i+}$ remains unchanged, in the octonion rotational transformations. It is easy to find that the octonion composite angular momentum, $( \textbf{i}_{\{L1i+\}} \circ \mathbb{L}^+ )$, in Eq.(7) is distinct from the term $\mathbb{L}^+$ in Eq.(5), or the term, $( \textbf{i}_{\{L1+\}} \circ \mathbb{L}^+ )$, in Eq.(6). Therefore the scalar parts of these three octonion composite angular momenta cannot be established simultaneously.

For the vectorial part, $\textbf{Z}_l^+ = ( i \textbf{L}_1^{i+} + \textbf{L}_1^+ ) + k_{eg} ( \textbf{L}_{20}^+ + i \textbf{L}_2^{i+} + \textbf{L}_2^+ )$ , of $\mathbb{L}^+$, if you multiply the unit vector, $\textbf{e}_{\{l+\}}$ , by $\textbf{Z}_l^+$ from the left, the scalar part, $\textbf{e}_{\{l+\}} \circ \textbf{Z}_l^+$ , will remain unchanged, in the rotational transformations of the octonion coordinate systems. Herein $\textbf{e}_{\{l+\}}$ is the unit vector of vectorial part $\textbf{Z}_l^+$ .

2) In the transformed octonion space $\mathbb{O}_u$ , if we multiply the unit vector, $i \textbf{I}_{\{L2+\}}$, by Eq.(5) from the left, it is able to achieve,
\begin{eqnarray}
i \textbf{I}_{\{L2+\}} \circ \mathbb{L}^+ = && i \textbf{I}_{\{L2+\}} \circ ( L_{10}^+ + i L_1^{i+} \textbf{i}_{\{L1i+\}} + L_1^+ \textbf{i}_{\{L1+\}} )
\nonumber
\\
&& + i k_{eg} ( \textbf{I}_{\{L2+\}} \circ \textbf{L}_{20}^+ + i L_2^{i+} \textbf{I}_{\{L2+\}} \circ \textbf{I}_{\{L2i+\}} - L_2^+  ) ~,
\end{eqnarray}
where $( \textbf{I}_{\{L2+\}} )^2 = -1$.

The scalar part, $( - L_2^+ )$, in the above will remain unchanged, in the rotational transformations of the octonion coordinate systems. In other words, the magnitude of magnetic moment will keep the same in the octonion rotational transformations. Apparently the octonion composite angular momentum, $( i \textbf{I}_{\{L2+\}} \circ \mathbb{L}^+ )$ , in Eq.(8) is different from that in Eqs.(5) to (7), respectively. As a result, the scalar parts of these four octonion composite angular momenta are unable to be established simultaneously. If the magnitude of magnetic moment is chosen as one conserved quantity, its application range will be restricted to some extent.

In the transformed octonion space $\mathbb{O}_u$ , if you multiply the unit vector, $i \textbf{I}_{\{L2i+\}}$, by Eq.(5) from the left, it is capable of attaining,
\begin{eqnarray}
i \textbf{I}_{\{L2i+\}} \circ \mathbb{L}^+ = && i \textbf{I}_{\{L2i+\}} \circ ( L_{10}^+ + i L_1^{i+} \textbf{i}_{\{L1i+\}} + L_1^+ \textbf{i}_{\{L1+\}} )
\nonumber
\\
&& + i k_{eg} ( \textbf{I}_{\{L2i+\}} \circ \textbf{L}_{20}^+ - i L_2^{i+} + L_2^+ \textbf{I}_{\{L2i+\}} \circ \textbf{I}_{\{L2+\}} ) ~,
\end{eqnarray}
where $( \textbf{I}_{\{L2i+\}} )^2 = -1$.

The scalar part, $( - i L_2^{i+} )$, in the above will remain unchanged, in the rotational transformations of the octonion coordinate systems. In other words, the magnitude of electric moment will keep the same in the octonion rotational transformations. Apparently the octonion composite angular momentum, $( i \textbf{I}_{\{L2i+\}} \circ \mathbb{L}^+ )$ , in Eq.(9) is distinct from that in Eqs.(5) to (8), respectively. Therefore the scalar parts of these five octonion composite angular momenta are unable to be established simultaneously. If the magnitude of electric moment is selected as one conserved quantity, its application range will be limited to some extent.

For the vectorial part, $\textbf{Z}_L^+$ , of transformed octonion composite angular momentum $i \textbf{I}_0 \circ \mathbb{L}^+$, if you multiply the unit vector, $\textbf{e}_{\{L+\}}$ , by $\textbf{Z}_L^+$ from the left, the scalar part, $\textbf{e}_{\{L+\}} \circ \textbf{Z}_L^+$ , will keep the same, in the rotational transformations of the octonion coordinate systems. Herein $\textbf{e}_{\{L+\}}$ is the unit vector of vectorial part $\textbf{Z}_L^+$ .

Because the transformed octonion space $\mathbb{O}_u$ is independent of the octonion space $\mathbb{O}$ , the magnitude of magnetic moment (or electric moment) and that of linear momentum cannot be selected as the conserved quantities simultaneously.

The above approach can be extended to some vectorial magnitudes of the octonion composite torque.

\begin{table}[h]
\tbl{Four conserved quantities, relevant to the vectorial magnitudes of the octonion torque, $\mathbb{W}^+$, for the electromagnetic and gravitational fields within the material media.}
{\begin{tabular}{@{}llll@{}}
\hline\hline
physical quantity               &  magnitude                      &  unit vector                   &  octonion space                \\
\hline
$\mathbb{W}^+$                  &  $W_1^{i+}$                     &  $\textbf{i}_{\{W1i+\}}$         &  $\mathbb{O}$                  \\
                                &  $W_1^+$                        &  $\textbf{i}_{\{W1+\}}$          &  $\mathbb{O}$                  \\
                                &  $W_2^{i+}$                     &  $\textbf{I}_{\{W2i+\}}$         &  $\mathbb{O}_u$                \\
                                &  $W_2^+$                        &  $\textbf{I}_{\{W2+\}}$          &  $\mathbb{O}_u$                \\
\hline\hline
\end{tabular}}
\end{table}

\section{\label{sec:level1}Composite torque}

By means of four unit vectors of composite torque in the octonion spaces, it is able to achieve four conserved quantities related to the vectorial magnitudes, under certain correlation conditions. a) The magnitude of torque is a conserved quantity; b) The magnitude of second-torque is a conserved quantity (Table 4).

The octonion composite torque can be defined as, $\mathbb{W}^+ = - v_0 ( i \mathbb{F}^+ / v_0 + \lozenge ) \circ \{ ( i \mathbb{V}^\times / v_0 ) \circ \mathbb{L}^+ \}$, within the material media. And that the formula can be rewritten as,
\begin{eqnarray}
\mathbb{W}^+ = && ( i W_{10}^{i+} + W_{10}^+ + i \textbf{W}_1^{i+} + \textbf{W}_1^+ )
\nonumber
\\
&&
+ k_{eg} ( i \textbf{W}_{20}^{i+} + \textbf{W}_{20}^+ + i \textbf{W}_2^{i+} + \textbf{W}_2^+ ) ~,
\end{eqnarray}
where $\textbf{W}_1^+ = \Sigma W_{1k}^+ \textbf{i}_k$ , $\textbf{W}_1^{i+} = \Sigma W_{1k}^{i+} \textbf{i}_k$ . $\textbf{W}_2^+ = \Sigma W_{2k}^+ \textbf{I}_k$ , $\textbf{W}_2^{i+} = \Sigma W_{2k}^{i+} \textbf{I}_k$ . $\textbf{W}_{20}^{i+} = W_{20}^{i+} \textbf{I}_0$ , $\textbf{W}_{20}^+ = W_{20}^+ \textbf{I}_0$ . $W_{10}^{i+}$ is the energy within the material media. $\textbf{W}_1^{i+}$ is the torque within the material media, including the gyroscopic torque. $\textbf{W}_2^{i+}$ is called as the second-torque within the material media temporarily. $W_{1j}^+$ , $W_{2j}^+$ , $W_{1j}^{i+}$, and $W_{2j}^{i+}$ are all real.

It is easy to find that the second torque relates with the derivative of magnetic moment and the curl of electric moment (see Ref.[28]). The second torque is in the 2-quaternion space $\mathbb{H}_e$ , and is similar to the torque in the quaternion space $\mathbb{H}_g$ .

The above can be further expanded as,
\begin{eqnarray}
\mathbb{W}^+ = && ( i W_{10}^{i+} + W_{10}^+ + i W_1^{i+} \textbf{i}_{\{W1i+\}} + W_1^+ \textbf{i}_{\{W1+\}} )
\nonumber
\\
&& + k_{eg} ( i \textbf{W}_{20}^{i+} + \textbf{W}_{20}^+ + i W_2^{i+} \textbf{I}_{\{W2i+\}} + W_2^+ \textbf{I}_{\{W2+\}} ) ~,
\end{eqnarray}
where each of $\textbf{i}_{\{W1+\}}$ , $\textbf{i}_{\{W1i+\}}$ , $\textbf{I}_{\{W2+\}}$ , and $\textbf{I}_{\{W2i+\}}$ is the unit vector. $\textbf{W}_1^{i+} = W_1^{i+} \textbf{i}_{\{W1i+\}}$ . $\textbf{W}_1^+ = W_1^+ \textbf{i}_{\{W1+\}}$ . $\textbf{W}_2^{i+} = W_2^{i+} \textbf{I}_{\{W2i+\}}$ . $\textbf{W}_2^+ = W_2^+ \textbf{I}_{\{W2+\}}$ .

In the octonion spaces, we had explored the conserved quantities relevant to the scalar parts, $W_{10}^+$ , $W_{20}^+$ , $W_{10}^{i+}$ and $W_{20}^{i+}$ , of the octonion composite torque (see Ref.[30]). In the paper, it is going to study the conserved quantities related to the vectorial parts, $\textbf{W}_1^{i+}$ , $\textbf{W}_1^+$ , $\textbf{W}_2^{i+}$ and $\textbf{W}_2^+$ , of the octonion composite torque.

1) In the octonion space $\mathbb{O}$ , if you multiply the unit vector, $\textbf{i}_{\{W1+\}}$ , by Eq.(11) from the left, it is going to attain,
\begin{eqnarray}
\textbf{i}_{\{W1+\}} \circ \mathbb{W}^+ = &&  ( i W_{10}^{i+} \textbf{i}_{\{W1+\}} + W_{10}^+ \textbf{i}_{\{W1+\}}
\nonumber
\\
&&
~~~~~~+ i W_1^{i+} \textbf{i}_{\{W1+\}} \circ \textbf{i}_{\{W1i+\}} - W_1^+ )
\nonumber
\\
&& + k_{eg} \textbf{i}_{\{W1+\}} \circ ( i \textbf{W}_{20}^{i+} + \textbf{W}_{20}^+
\nonumber
\\
&&
~~~~~~+ i W_2^{i+} \textbf{I}_{\{W2i+\}} + W_2^+ \textbf{I}_{\{W2+\}} ) ~,
\end{eqnarray}
where $( \textbf{i}_{\{W1+\}} )^2 = -1$.

The scalar part, $( - W_1^+ )$, in the above remains unchanged, in the rotational transformations of the octonion coordinate systems. In other words, the magnitude, $W_1^+$ , will keep the same in the octonion rotational transformations. It should be noted that the octonion composite torque, $( \textbf{i}_{\{W1+\}} \circ \mathbb{W}^+ )$, in Eq.(12) is distinct from $\mathbb{W}^+$ in Eq.(11). It means that the scalar parts of these two octonion composite torques cannot be effective simultaneously.

In the octonion space $\mathbb{O}$ , if we multiply the unit vector, $\textbf{i}_{\{W1i+\}}$ , by Eq.(11) from the left, it is able to obtain,
\begin{eqnarray}
\textbf{i}_{\{W1i+\}} \circ \mathbb{W}^+ = && ( i W_{10}^{i+} \textbf{i}_{\{W1i+\}}  + W_{10}^+ \textbf{i}_{\{W1i+\}}
\nonumber
\\
&&
~~~~~~- i W_1^{i+} + W_1^+ \textbf{i}_{\{W1i+\}} \circ \textbf{i}_{\{W1+\}} )
\nonumber
\\
&& + k_{eg} \textbf{i}_{\{W1i+\}} \circ ( i \textbf{W}_{20}^{i+} + \textbf{W}_{20}^+
\nonumber
\\
&&
~~~~~~+ i W_2^{i+} \textbf{I}_{\{W2i+\}} + W_2^+ \textbf{I}_{\{W2+\}} ) ~,
\end{eqnarray}
where $( \textbf{i}_{\{W1i+\}} )^2 = -1$.

In the rotational transformations of the octonion coordinate systems, the scalar part, $( - i W_1^{i+} )$, in the above keeps the same. In other words, the magnitude of torque remains unchanged, in the octonion rotational transformations. Apparently the octonion composite torque, $( \textbf{i}_{\{W1i+\}} \circ \mathbb{W}^+ )$, in Eq.(13) is distinct from that in Eq.(11) or Eq.(12). Therefore the scalar parts of these three octonion composite torques cannot be established simultaneously. If the magnitude of torque is selected as one conserved quantity, its application range will be restrained to some extent.

For the vectorial part, $\textbf{Z}_w^+$ , of $\mathbb{W}^+$, if you multiply the unit vector, $\textbf{e}_{\{w+\}}$ , by $\textbf{Z}_w^+$ from the left, the scalar part, $\textbf{e}_{\{w+\}} \circ \textbf{Z}_w^+$ , will remain unchanged, in the rotational transformations of the octonion coordinate systems. Herein $\textbf{e}_{\{w+\}}$ is the unit vector of vectorial part $\textbf{Z}_w^+$ .

2) In the transformed octonion space $\mathbb{O}_u$ , if you multiply the unit vector, $i \textbf{I}_{\{W2+\}}$, by Eq.(11) from the left, it is able to achieve,
\begin{eqnarray}
i \textbf{I}_{\{W2+\}} \circ \mathbb{W}^+ = && i \textbf{I}_{\{W2+\}} \circ ( i W_{10}^{i+} + W_{10}^+
\nonumber
\\
&&
~~~~~~+ i W_1^{i+} \textbf{i}_{\{W1i+\}} + W_1^+ \textbf{i}_{\{W1+\}} )
\nonumber
\\
&& + i k_{eg} ( i \textbf{I}_{\{W2+\}} \circ \textbf{W}_{20}^{i+} + \textbf{I}_{\{W2+\}} \circ \textbf{W}_{20}^+
\nonumber
\\
&&
~~~~~~+ i W_2^{i+} \textbf{I}_{\{W2+\}} \circ \textbf{I}_{\{W2i+\}} - W_2^+ ) ~,
\end{eqnarray}
where $( \textbf{I}_{\{W2+\}} )^2 = -1$.

The scalar part, $( - W_2^+ )$, in the above will remain unchanged, in the rotational transformations of the octonion coordinate systems. In other words, the magnitude, $W_2^+$ , will keep the same in the octonion rotational transformations. Apparently the octonion composite torque, $( i \textbf{I}_{\{W2+\}} \circ \mathbb{W}^+ )$ , in Eq.(14) is different from that in Eqs.(11) to (13). As a result, the scalar parts of these four octonion composite torques are unable to be established simultaneously.

In the transformed octonion space $\mathbb{O}_u$ , if we multiply the unit vector, $i \textbf{I}_{\{W2i+\}}$, by Eq.(11) from the left, it is able to attain,
\begin{eqnarray}
i \textbf{I}_{\{W2i+\}} \circ \mathbb{W}^+ = && i \textbf{I}_{\{W2i+\}} \circ ( i W_{10}^{i+} + W_{10}^+
\nonumber
\\
&&
~~~~~~+ i W_1^{i+} \textbf{i}_{\{W1i+\}} + W_1^+ \textbf{i}_{\{W1+\}} )
\nonumber
\\
&& + i k_{eg} ( i \textbf{I}_{\{W2i+\}} \circ \textbf{W}_{20}^{i+} + \textbf{I}_{\{W2i+\}} \circ \textbf{W}_{20}^+
\nonumber
\\
&&
~~~~~~- i W_2^{i+} + W_2^+  \textbf{I}_{\{W2i+\}} \circ \textbf{I}_{\{W2+\}} ) ~,
\end{eqnarray}
where $( \textbf{I}_{\{W2i+\}} )^2 = -1$.

The scalar part, $( - i W_2^{i+} )$, in the above will remain unchanged, in the rotational transformations of the octonion coordinate systems. In other words, the magnitude of second-torque will keep the same in the octonion rotational transformations. Apparently the octonion composite torque, $( i \textbf{I}_{\{W2i+\}} \circ \mathbb{W}^+ )$, in Eq.(15) is distinct from those in Eqs.(11) to (14). Consequently the scalar parts of these five octonion composite torques are unable to be established simultaneously.

For the vectorial part, $\textbf{Z}_W^+$, of transformed octonion composite torque $i \textbf{I}_0 \circ \mathbb{W}^+$, if you multiply the unit vector, $\textbf{e}_{\{W+\}}$, by $\textbf{Z}_W^+$ from the left, the scalar part, $\textbf{e}_{\{W+\}} \circ \textbf{Z}_W^+$, will remain unchanged, in the rotational transformations of the octonion coordinate systems. Herein $\textbf{e}_{\{W+\}}$ is the unit vector of vectorial part $\textbf{Z}_W^+$.

Since the transformed octonion space $\mathbb{O}_u$ is independent of the octonion space $\mathbb{O}$ , the magnitude of torque and that of second-torque cannot be selected as the conserved quantities simultaneously.

The above approach can be extended to some vectorial magnitudes of the octonion composite force.

\begin{table}[h]
\tbl{Four conserved quantities, relevant to the vectorial magnitudes of the octonion force, $\mathbb{N}^+$, for the electromagnetic and gravitational fields within the material media.}
{\begin{tabular}{@{}llll@{}}
\hline\hline
physical quantity               &  magnitude                      &  unit vector                   &  octonion space                \\
\hline
$\mathbb{N}^+$                  &  $N_1^{i+}$                     &  $\textbf{i}_{\{N1i+\}}$         &  $\mathbb{O}$                  \\
                                &  $N_1^+$                        &  $\textbf{i}_{\{N1+\}}$          &  $\mathbb{O}$                  \\
                                &  $N_2^{i+}$                     &  $\textbf{I}_{\{N2i+\}}$         &  $\mathbb{O}_u$                \\
                                &  $N_2^+$                        &  $\textbf{I}_{\{N2+\}}$          &  $\mathbb{O}_u$                \\
\hline\hline
\end{tabular}}
\end{table}

\section{\label{sec:level1}Composite force}

Making use of four unit vectors of the composite force in the octonion spaces, it is able to achieve four conserved quantities related to the vectorial magnitudes, under certain correlation conditions. a) The magnitude of force is a conserved quantity; b) The magnitude of second-force is a conserved quantity (Table 5).

The octonion composite force can be defined as, $\mathbb{N}^+ = - ( i \mathbb{F}^+ / v_0 + \lozenge ) \circ \{ ( i \mathbb{V}^\times / v_0 ) \circ \mathbb{W}^+ \}$, within the material media. And that the formula can be rewritten as,
\begin{eqnarray}
\mathbb{N}^+ = && ( i N_{10}^{i+} + N_{10}^+ + i \textbf{N}_1^{i+} + \textbf{N}_1^+ )
\nonumber
\\
&&
+ k_{eg} ( i \textbf{N}_{20}^{i+} + \textbf{N}_{20}^+ + i \textbf{N}_2^{i+} + \textbf{N}_2^+ ) ~,
\end{eqnarray}
where $\textbf{N}_1^+ = \Sigma N_{1k}^+ \textbf{i}_k$ , $\textbf{N}_1^{i+} = \Sigma N_{1k}^{i+} \textbf{i}_k$ . $\textbf{N}_2^+ = \Sigma N_{2k}^+ \textbf{I}_k$ , $\textbf{N}_2^{i+} = \Sigma N_{2k}^{i+} \textbf{I}_k$ . $\textbf{N}_{20}^{i+} = N_{20}^{i+} \textbf{I}_0$ , $\textbf{N}_{20}^+ = N_{20}^+ \textbf{I}_0$ . $N_{10}^+$ is the power within the material media. $\textbf{N}_1^{i+}$ is the force within the material media, including the Magnus force. $\textbf{N}_2^{i+}$ is called as the second-force within the material media temporarily. $N_{1j}^+$ , $N_{2j}^+$ , $N_{1j}^{i+}$ , and $N_{2j}^{i+}$ are all real.

One can find that the second force relates with the derivative of `the curl of magnetic moment and the derivative of electric moment' (see Ref.[28]). The second force is in the 2-quaternion space $\mathbb{H}_e$ , and is similar to the force in the quaternion space $\mathbb{H}_g$ .

The above can be further expanded as,
\begin{eqnarray}
\mathbb{N}^+ = && ( i N_{10}^{i+} + N_{10}^+ + i N_1^{i+} \textbf{i}_{\{N1i+\}} + N_1^+ \textbf{i}_{\{N1+\}} )
\nonumber
\\
&& + k_{eg} ( i \textbf{N}_{20}^{i+} + \textbf{N}_{20}^+ + i N_2^{i+} \textbf{I}_{\{N2i+\}} + N_2^+ \textbf{I}_{\{N2+\}} ) ~,
\end{eqnarray}
where each of $\textbf{i}_{\{N1+\}}$ , $\textbf{i}_{\{N1i+\}}$ , $\textbf{I}_{\{N2+\}}$ , and $\textbf{I}_{\{N2i+\}}$ is the unit vector. $\textbf{N}_1^{i+} = N_1^{i+} \textbf{i}_{\{N1i+\}}$ . $\textbf{N}_1^+ = N_1^+ \textbf{i}_{\{N1+\}}$ . $\textbf{N}_2^{i+} = N_2^{i+} \textbf{I}_{\{N2i+\}}$ . $\textbf{N}_2^+ = N_2^+ \textbf{I}_{\{N2+\}}$ .

In the octonion spaces, we had discussed the conserved quantities relevant to the scalar parts, $N_{10}^+$ , $N_{20}^+$ , $N_{10}^{i+}$ and $N_{20}^{i+}$ , of the octonion composite force (see Ref.[30]). In the paper, it is going to study the conserved quantities related to the vectorial parts, $\textbf{N}_1^{i+}$ , $\textbf{N}_1^+$ , $\textbf{N}_2^{i+}$ and $\textbf{N}_2^+$ , of the octonion composite force.

1) In the octonion space $\mathbb{O}$ , if you multiply the unit vector, $\textbf{i}_{\{N1+\}}$ , by Eq.(17) from the left, it is going to get,
\begin{eqnarray}
\textbf{i}_{\{N1+\}} \circ \mathbb{N}^+ = && ( i N_{10}^{i+} \textbf{i}_{\{N1+\}} + N_{10}^+ \textbf{i}_{\{N1+\}}
\nonumber
\\
&&
~~~~~~+ i N_1^{i+} \textbf{i}_{\{N1+\}} \circ \textbf{i}_{\{N1i+\}} - N_1^+ )
\nonumber
\\
&& + k_{eg} \textbf{i}_{\{N1+\}} \circ ( i \textbf{N}_{20}^{i+} + \textbf{N}_{20}^+
\nonumber
\\
&&
~~~~~~+ i N_2^{i+} \textbf{I}_{\{N2i+\}} + N_2^+ \textbf{I}_{\{N2+\}} ) ~,
\end{eqnarray}
where $( \textbf{i}_{\{N1+\}} )^2 = -1$.

The scalar part, $( - N_1^+ )$, in the above remains unchanged, in the rotational transformations of the octonion coordinate systems. In other words, the magnitude, $N_1^+$ , will keep the same in the octonion rotational transformations. Significantly, the octonion composite force, $( \textbf{i}_{\{N1+\}} \circ \mathbb{N}^+ )$, in Eq.(18) is distinct from $\mathbb{N}^+$ in Eq.(17). It means that the scalar parts of these two octonion composite forces cannot be effective simultaneously.

In the octonion space $\mathbb{O}$ , if we multiply the unit vector, $\textbf{i}_{\{N1i+\}}$ , by Eq.(17) from the left, it is able to achieve,
\begin{eqnarray}
\textbf{i}_{\{N1i+\}} \circ \mathbb{N}^+ = && ( i N_{10}^{i+} \textbf{i}_{\{N1i+\}} + N_{10}^+ \textbf{i}_{\{N1i+\}}
\nonumber
\\
&&
~~~~~~- i N_1^{i+} + N_1^+ \textbf{i}_{\{N1i+\}} \circ \textbf{i}_{\{N1+\}} )
\nonumber
\\
&& + k_{eg} \textbf{i}_{\{N1i+\}} \circ ( i \textbf{N}_{20}^{i+} + \textbf{N}_{20}^+
\nonumber
\\
&&
~~~~~~+ i N_2^{i+} \textbf{I}_{\{N2i+\}} + N_2^+ \textbf{I}_{\{N2+\}} ) ~,
\end{eqnarray}
where $( \textbf{i}_{\{N1i+\}} )^2 = -1$.

In the rotational transformations of the octonion coordinate systems, the scalar part, $( - i N_1^{i+} )$, in the above keeps the same. In other words, the magnitude of force remains unchanged, in the octonion rotational transformations. Obviously the octonion composite force, $( \textbf{i}_{\{N1i+\}} \circ \mathbb{N}^+ )$, in Eq.(19) is distinct from that in Eq.(17) or Eq.(18). Therefore the scalar parts of these three octonion composite forces cannot be established simultaneously. If the magnitude of force is chosen as one conserved quantity, its application range will be confined to some extent.

For the vectorial part, $\textbf{Z}_n^+$ , of $\mathbb{N}^+$, if you multiply the unit vector, $\textbf{e}_{\{n+\}}$ , by $\textbf{Z}_n^+$ from the left, the scalar part, $\textbf{e}_{\{n+\}} \circ \textbf{Z}_n^+$ , will keep the same, in the rotational transformations of the octonion coordinate systems. Herein $\textbf{e}_{\{n+\}}$ is the unit vector of vectorial part $\textbf{Z}_n^+$ .

2) In the transformed octonion space $\mathbb{O}_u$ , if you multiply the unit vector, $i \textbf{I}_{\{N2+\}}$, by Eq.(17) from the left, it is going to achieve,
\begin{eqnarray}
i \textbf{I}_{\{N2+\}} \circ \mathbb{N}^+ = && i \textbf{I}_{\{N2+\}} \circ ( i N_{10}^{i+} + N_{10}^+
\nonumber
\\
&&
~~~~~~+ i N_1^{i+} \textbf{i}_{\{N1i+\}} + N_1^+ \textbf{i}_{\{N1+\}} )
\nonumber
\\
&& + i k_{eg} ( i \textbf{I}_{\{N2+\}} \circ \textbf{N}_{20}^{i+} + \textbf{I}_{\{N2+\}} \circ \textbf{N}_{20}^+
\nonumber
\\
&&
~~~~~~+ i N_2^{i+} \textbf{I}_{\{N2+\}} \circ \textbf{I}_{\{N2i+\}} - N_2^+ ) ~,
\end{eqnarray}
where $( \textbf{I}_{\{N2+\}} )^2 = -1$.

The scalar part, $( - N_2^+ )$, in the above will remain unchanged, in the rotational transformations of the octonion coordinate systems. In other words, the magnitude, $N_2^+$ , will keep the same in the octonion rotational transformations. Apparently the octonion composite force, $( i \textbf{I}_{\{N2+\}} \circ \mathbb{N}^+ )$ , in Eq.(20) is different from those in Eqs.(17) to (19). As a result, the scalar parts of these four octonion composite forces are incompatible  simultaneously.

In the transformed octonion space $\mathbb{O}_u$ , if we multiply the unit vector, $i \textbf{I}_{\{N2i+\}}$, by Eq.(17) from the left, we shall attain,
\begin{eqnarray}
i \textbf{I}_{\{N2i+\}} \circ \mathbb{N}^+ = && i \textbf{I}_{\{N2i+\}} \circ ( i N_{10}^{i+} + N_{10}^+
\nonumber
\\
&&
~~~~~~+ i N_1^{i+} \textbf{i}_{\{N1i+\}} + N_1^+ \textbf{i}_{\{N1+\}} )
\nonumber
\\
&& + i k_{eg} ( i \textbf{I}_{\{N2i+\}} \circ \textbf{N}_{20}^{i+} + \textbf{I}_{\{N2i+\}} \circ \textbf{N}_{20}^+
\nonumber
\\
&&
~~~~~~- i N_2^{i+} + N_2^+ \textbf{I}_{\{N2i+\}} \circ \textbf{I}_{\{N2+\}} ) ~,
\end{eqnarray}
where $( \textbf{I}_{\{N2i+\}} )^2 = -1$.

The scalar part, $( - i N_2^{i+} )$, in the above will remain unchanged, in the rotational transformations of the octonion coordinate systems. In other words, the magnitude of second-force will keep the same in the octonion rotational transformations. Apparently the octonion composite force, $( i \textbf{I}_{\{N2i+\}} \circ \mathbb{N}^+ )$, in Eq.(21) is distinct from those in Eqs.(17) to (20). Consequently the scalar parts of these five octonion composite forces are unable to be established simultaneously.

For the vectorial part, $\textbf{Z}_N^+$ , of transformed octonion composite force $i \textbf{I}_0 \circ \mathbb{N}^+$, if you multiply the unit vector, $\textbf{e}_{\{N+\}}$ , by $\textbf{Z}_N^+$ from the left, the scalar part, $\textbf{e}_{\{N+\}} \circ \textbf{Z}_N^+$, will keep the same, in the rotational transformations of the octonion coordinate systems. Herein $\textbf{e}_{\{N+\}}$ is the unit vector of vectorial part $\textbf{Z}_N^+$ .

Since the transformed octonion space $\mathbb{O}_u$ is independent of the octonion space $\mathbb{O}$, the magnitude of force and that of second-force cannot be selected as the conserved quantities simultaneously.

To sum up, the case where the vectorial magnitude is taken as the conserved quantity will be subject to more restrictions, in the octonion spaces. Its scope of application will be comparatively narrow.

\begin{table}[h]
\tbl{Some conserved quantities related to vectorial magnitudes within the material media, in the octonion spaces for the electromagnetic and gravitational fields.}
{\begin{tabular}{@{}lll@{}}
\hline\hline
physical quantity    &  octonion space $\mathbb{O}$                      &  transformed octonion space $\mathbb{O}_u$                             \\
\hline
$\mathbb{P}^+$       &  $p^+$ , magnitude of linear momentum             &  $P^+$ , magnitude of electric current                                 \\
$\mathbb{L}^+$       &  $L_1^{i+}$                                       &  $L_2^{i+}$ , magnitude of electric moment                             \\
                     &  $L_1^+$ , magnitude of angular momentum          &  $L_2^+$ , magnitude of magnetic moment                                \\
$\mathbb{W}^+$       &  $W_1^{i+}$ , magnitude of torque                 &  $W_2^{i+}$ , magnitude of second-torque                               \\
                     &  $W_1^+$                                          &  $W_2^+$                                                               \\
$\mathbb{N}^+$       &  $N_1^{i+}$ , magnitude of force                  &  $N_2^{i+}$ , magnitude of second-force                                \\
                     &  $N_1^+$                                          &  $N_2^+$                                                               \\
\hline\hline
\end{tabular}}
\end{table}

\section{\label{sec:level1}Discussions and conclusions}

The application of octonion spaces can describe simultaneously the physical quantities of gravitational and electromagnetic fields, including the octonion field strength, field source, linear momentum, angular momentum, torque and force within the material media. In the rotational transformation of the octonion coordinate systems, the scalar part of one octonion physical quantity remains unchanged, although the vectorial part is variable. Some conserved quantities can be derived from the property of octonions. Especially, it is able to achieve some conserved quantities related to the magnitudes of vectors, by means of the characteristics of octonion rotational transformations (Table 6).

In the gravitational and electromagnetic field within the material media, there is a problem of simultaneity among the conserved quantities, according to the octonion field theory. Since the octonion space $\mathbb{O}$ is independent of the transformed octonion space $\mathbb{O}_u$ , the conserved quantities in the octonion space $\mathbb{O}$ are unable to be established simultaneously with the conserved quantities in the transformed octonion space $\mathbb{O}_u$ . Furthermore the conserved quantities of some octonion physical quantities cannot be effective simultaneously, for the octonion space $\mathbb{O}$ (or transformed octonion space $\mathbb{O}_u$ ). In the physics, the conserved quantities related to the magnitudes of vectors are applied to a great extent. However, the simultaneity of conserved quantities place more restrictions on the applicable conditions of the conserved quantities relevant to the magnitudes of vectors, compared with the conserved quantities related to the scalar parts of octonion physical quantities. It means that the scope of application of some conserved quantities related to vectorial magnitudes is extremely limited, due to the restrictions of simultaneity.

Some classical conserved quantities are unable to be established simultaneously, in the octonion spaces within the material media. a) The magnitude of linear momentum and that of electric current cannot be simultaneously selected as the conserved quantities within the material media, in terms of the octonion composite linear momentum. b) The magnitudes of electric moment, magnetic moment and angular momentum cannot be simultaneously chosen as the conserved quantities within the material media, for the octonion composite angular momentum. c) The magnitude of torque and that of second-torque will not be simultaneously selected as the conserved quantities within the material media, in terms of the octonion composite torque. d) The magnitudes of force and second-force may not be simultaneously chosen as the conserved quantities within the material media, for the octonion composite force. This result enriches our understanding of the physical properties of conserved quantities.

According to the octonion field theory, the charge-to-mass ratio of a charged particle will certainly vary with the velocity and other factors. Since the octonion space $\mathbb{O}$ is independent of the transformed octonion space $\mathbb{O}_u$ , the mass and the quantity of electric charge are unable to be selected as the conserved quantities simultaneously. Similarly, the magnitude of linear momentum and that of electric current cannot be chosen as the conserved quantities simultaneously. Either two magnitudes of angular momentum and magnetic moment cannot be selected as the conserved quantities simultaneously, in the strict sense. These factors are some of important reasons for the variation of charge-to-mass ratio of charged particles.

For the uniform relative motion between two coordinate systems, the coordinate transformation in the four-dimensional spaces is different from that in the two-dimensional spaces. The tensor analysis can be utilized to study some characteristics of coordinate transformations, in the special theory of relativity related to the four-dimensional spacetime. The coordinate transformation between two coordinate systems of one uniform relative motion, in the two-dimensional spacetime, is actually merely the rotational transformation in the four-dimensional spacetime. Similarly, the coordinate transformation between two coordinate systems of one uniform relative motion, in the two-dimensional space of one quaternion space, is actually only the rotational transformation in the quaternion space, according to the algebra of quaternions. Furthermore, this method can also be extended to the related cases in the octonion spaces.

It is worth noting that the paper just discusses some conserved quantities and their simultaneity related to the magnitudes of vectors in the octonion spaces. But it has been clearly explained that the scalar part of one octonion physical quantity can remain unchanged, in the rotational transformations of the octonion coordinate systems. By means of this octonion property, it is able to deduce some conserved quantities relevant to the magnitudes of vectors. Furthermore some conserved quantities related to the magnitudes of vectors cannot be established simultaneously, according to the algebra of octonions. In the future study, it is going to further research other conserved quantities related to the magnitudes of vectors, exploring some conditions under which the conserved quantities can be established simultaneously.

\section*{Acknowledgments}
The author is indebted to the anonymous referees for their valuable comments on the previous manuscripts. This project was supported partially by the National Natural Science Foundation of China under grant number 60677039.


\end{document}